\documentclass[%
 reprint,
 amsmath,amssymb,
 aps,
]{revtex4-2}

\usepackage{graphicx}
\usepackage{dcolumn}
\usepackage{bm}
\usepackage{braket}
\usepackage{mathrsfs}  
\usepackage{xfrac}
\renewcommand\bra[1]{{\langle{#1}|}}
\makeatletter
\renewcommand\ket[1]{%
  \@ifnextchar\bra{\k@t{#1}\!}{\k@t{#1}}%
}
\newcommand\k@t[1]{{|{#1}\rangle}}
\makeatother

\begin{document}

\preprint{APS/123-QED}

\title{Tight information bounds for spontaneous emission lifetime resolution of quantum sources with varied spectral purity}

\author{Cheyenne S. Mitchell}
\author{Mikael P. Backlund}
\email{mikaelb@illinois.edu}
\affiliation{%
 Department of Chemistry and Illinois Quantum Information Science and Technology Center (IQUIST), University of Illinois at Urbana-Champaign, Urbana, IL 61801
}%

\date{\today}

\begin{abstract}
We generalize the theory of resolving a mixture of two closely spaced spontaneous emission lifetimes to include pure dephasing contributions to decoherence, leading to the resurgence of Rayleigh's Curse at small lifetime separations. Considerable resolution enhancement remains possible when lifetime broadening is more significant than that due to pure dephasing. In the limit that lifetime broadening dominates, one can achieve super-resolution either by a tailored one-photon measurement or Hong-Ou-Mandel interferometry. We describe conditions for which either choice is superior. 
\end{abstract}

\maketitle


\section{\label{section_introduction}Introduction\\}
 A recent quantum-inspired analysis of the age-old problem of spatially resolving mutually incoherent optical point sources revealed conditions under which the precision of such a measurement can surpass Rayleigh's Curse \cite{tsang2016quantum,doi:10.1080/00107514.2020.1736375}. The authors showed that the quantum Fisher information (QFI) associated with estimation of the separation between two such point sources remains constant as the separation goes to zero, even as the classical Fisher information (CFI) associated with direct imaging vanishes in the same limit. A flurry of subsequent studies have since built on the theory \cite{PhysRevLett.117.190801,Nair:16,Tsang_2017,PhysRevA.95.063847,PhysRevA.97.023830,Rehacek:17,PhysRevA.99.012305,PhysRevA.99.013808,Bisketzi_2019,PhysRevLett.124.080503,Bao:21,bojer2022quantitative,Datta:21,GracePRL2022,HuangPRL2021,HuangPRA2023,Jusuf:22,KaruseichykPRR2022,krovi2022superresolution,KurdzialekQuantum2022,Hradil:19,LiangPRA2021,SorelliPRA2021,PhysRevA.104.022613} and experimentally demonstrated advantages in model imaging systems \cite{Yang:16,Tang:16,Boucher:20,Wadood:21,Zhou:19,PhysRevLett.118.070801,Paur:16,Paur:18,santamaria2023spatialmodedemultiplexing,Zhang:20}. The basic idea has also been translated from position-momentum to time-frequency resolution \cite{PhysRevLett.121.090501,PhysRevResearch.3.033082,PRXQuantum.2.010301,ShahPRApplied2021}. In this spirit, we recently reported quantum limits associated with the estimation, resolution, and discrimination of optical spontaneous emission lifetimes \cite{MitchellPRA2022}.

An important contingent of this body of research has presented caveats to the theory that effectively temper one's ability to surpass Rayleigh's Curse subject to certain experimental realities. Quantitatively, these caveats cause the QFI to eventually scale to zero as the separation becomes sufficiently small. Imperfect knowledge of the centroid position of the two sources is one such caveat that was explicitly noted from the beginning \cite{tsang2016quantum,Grace:20}. A mode-sorting measurement that would otherwise saturate the bound yields equivalent trends under misalignment or in the presence of crosstalk \cite{GessnerPRL2020,SorelliPRL2021,AlmeidaPRA2021}. Various other nuisance parameters \cite{RehacekPRA2017,Larson:18,Liang:21,linowski2022application,TanIEEE2022,Wang:21} or additional sources of noise \cite{kurdzialek2023measurement,len2020resolution,LupoPRA2020,OhPRL2021} can lead to similar mitigation. In this letter, we detail another such caveat that is specifically relevant to the resolution of optical spontaneous emission lifetime mixtures. Namely, diminished spectral purity of the collected photons leads to a lowering of the associated QFI, eventually recovering the classical bound associated with direct measurement via time-correlated single-photon counting (TCSPC). We quantify the relation between spectral purity and QFI for this system, and show that a significant resolution enhancement remains possible in the case that lifetime broadening is more significant than broadening due to pure dephasing. In the limit of high spectral purity we consider the prospect of attaining super-resolution via Hong-Ou-Mandel (HOM) interference  measurements \cite{PhysRevLett.59.2044} on subsequently emitted photons and compare performance to a tailored one-photon measurement scheme.
\section{\label{section_results}Results and Discussion\\}
For context, in Ref. \cite{MitchellPRA2022} we considered the task of estimating constituent lifetimes $\tau_0$ and $\tau_1$ given a mixed single-photon state of the form:
\begin{equation} \label{eq_rho_mixed_old}
    \rho = \frac{1}{2}\ket{\psi_{\tau_0}(\omega)}\bra{\psi_{\tau_0}(\omega)} + \frac{1}{2}\ket{\psi_{\tau_1}(\omega)}\bra{\psi_{\tau_1}(\omega)},
\end{equation}
wherein
\begin{equation}
    \ket{\psi_{\tau}(\omega)} = \int \mathrm{d}t \, \psi_{\tau}(t;\omega) a^\dagger(t) \ket{0} = \int \mathrm{d}t \, \psi_\tau(t;\omega) \ket{t},
\end{equation}
with $a^\dagger(t)$ the creation operator for the denoted temporal mode and
\begin{equation} \label{eq_onephotonwf}
    \psi_{\tau}(t;\omega) = \frac{H(t)}{\sqrt{\tau}}e^{-i \omega t}e^{-t/2\tau}
\end{equation}
for $\tau \in \{\tau_0, \tau_1\}$. In Eq. (\ref{eq_onephotonwf}) $H(t)$ is the Heaviside step function defined by $H(t\geq0) = 1$ and $H(t<0) = 0$. Definition of time $t$ is shifted to compensate for the finite distance between emitter and detector, such that the time window of interest begins at $t=0$. We found that the conventional measurement scheme based on TCSPC suffers from an analogue of Rayleigh's Curse in that the CFI associated with estimation of the square-root-ratio of lifetimes $\varepsilon = \sqrt{\tau_1/\tau_0}$ vanishes in the limit $\varepsilon \to 1$. By contrast, the QFI associated with estimating $\varepsilon$ attains its maximum in the same limit. We showed that this quantum bound is saturated by a projective measurement onto the basis of weighted Laguerre (WL) modes defined by:
\begin{equation} \label{eq_WLket}
    \ket{\phi_n(\omega,\bar{\tau})} = \int \mathrm{d}t \, \phi_n(t;\omega,\bar{\tau}) \ket{t}
\end{equation}
with \begin{equation} \label{eq_WLwf}
    \phi_n(t;\omega,\bar{\tau}) = \frac{H(t)}{\sqrt{\bar{\tau}}}e^{-i \omega t}e^{-t/2\bar{\tau}}L_n(t/\bar{\tau}),
\end{equation}
where $\bar{\tau} = \sqrt{\tau_0 \tau_1}$ is the geometric mean lifetime and $L_n(\cdot)$ denotes the Laguerre polynomial of order $n$. We considered possible routes to experimental realization of such a measurement as well as approximate interferometric schemes that outperform TCSPC.
\begin{figure}
    \centering
    \includegraphics{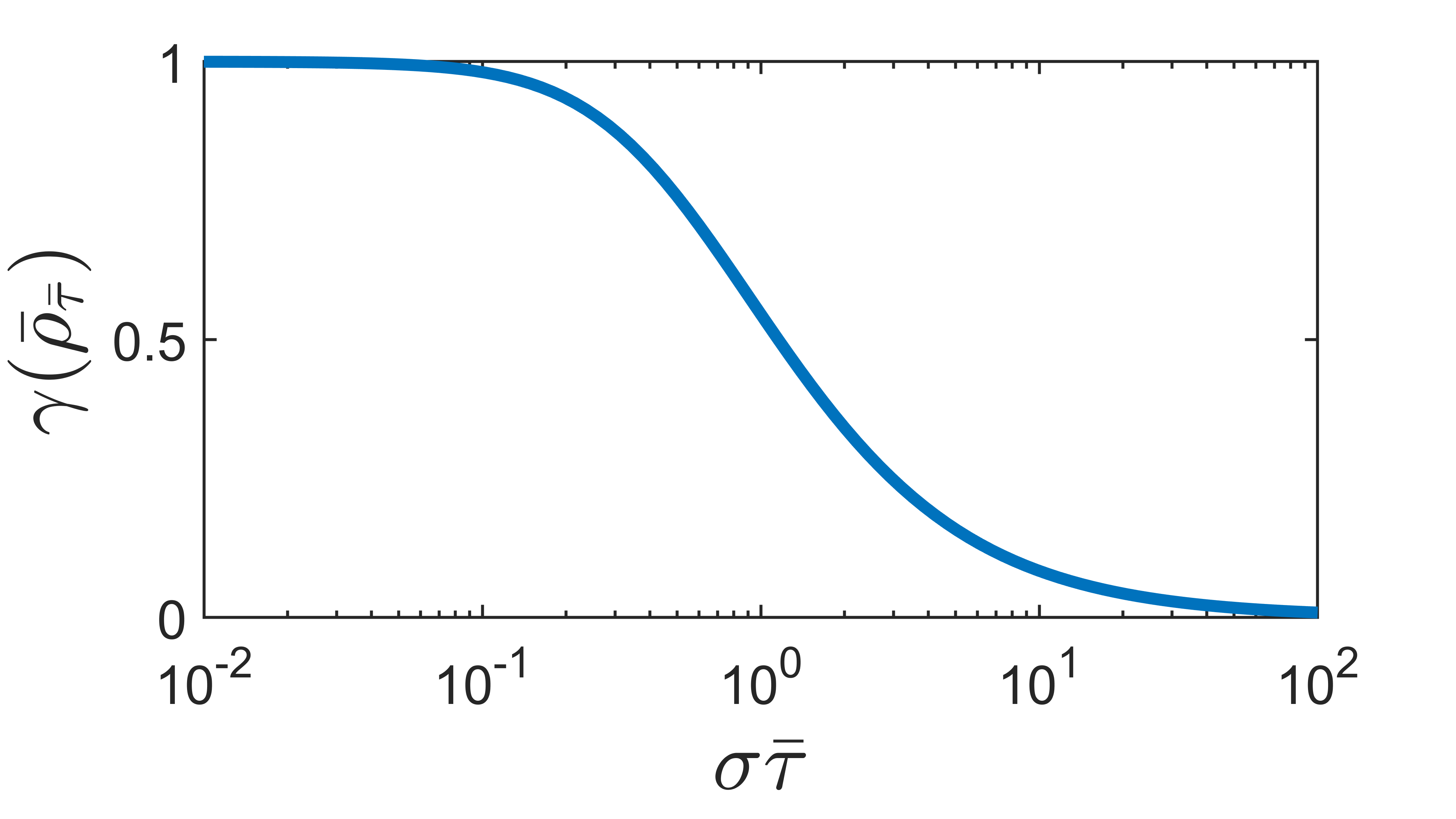}
    \caption{Purity of the limiting state $\bar{\rho}_{\bar{\tau}}$ as a function of the product $\sigma \bar{\tau}$.}
    \label{fig:purity}
\end{figure}
Though it provides a useful starting point, the model underlying Eqs. (\ref{eq_rho_mixed_old}-\ref{eq_onephotonwf}) employs several simplifying suppositions. In the current letter we will focus on one of these suppositions in particular: that the constituent single-lifetime states $\rho_{\tau_0}$ and $\rho_{\tau_1}$ are of unit purity, corresponding to photons whose spectral linewidths are lifetime-limited. In realistic systems one must contend with (often dominant) pure dephasing contributions to the spectral linewidth due to inhomogenous broadening (for emitter ensembles) and/or spectral diffusion (for single emitters). One typically has to work hard to produce lifetime-limited photons, either by freezing out sources of dephasing \cite{aharonovich2016solid} or engineering accelerated emission rates \cite{BogdanovScience2019}. 

Here we amend our model such that the collected single-photon state is given by:
\begin{equation}
    \bar{\rho} = \frac{1}{2}\left( \bar{\rho}_{\tau_0} + \bar{\rho}_{\tau_1} \right),
\end{equation}
where the overbar denotes incoherent averaging over a spectral density function $P(\omega)$ such that
\begin{equation}
    \bar{\rho}_{\tau} = \int \mathrm{d}\omega P(\omega) \ket{\psi_\tau(\omega)}\bra{\psi_\tau(\omega)}
\end{equation}
for $\tau \in \{\tau_0,\tau_1\}$. To isolate the resolution problem, we assume $\bar{\tau}$ is known and set out to calculate $\mathcal{K}_\varepsilon$, the QFI associated with $\varepsilon$, for various choices of $P(\omega)$. We take $P(\omega)$ to be centered about some known frequency $\omega_0 > 0$ such that $P(\omega) = P_0(\omega-\omega_0)$, where $P_0(\omega)$ is centered at $\omega = 0$. The QFI is given by
\begin{equation}
    \mathcal{K}_\varepsilon = \text{Tr}\left( \mathcal{L}_\varepsilon^2 \bar{\rho} \right),
\end{equation}
where $\mathcal{L}_\varepsilon$ is the symmetric logarithmic derivative (SLD) operator defined implicitly via
\begin{equation}
    \partial_\varepsilon \bar{\rho} = \frac{1}{2}(\mathcal{L}_\varepsilon \bar{\rho} + \bar{\rho} \mathcal{L}_\varepsilon).
\end{equation}
The SLD can be computed explicitly by first diagonalizing $\bar{\rho}$ such that
\begin{equation}
    \bar{\rho} = \sum_k D_k \ket{k}\bra{k}
\end{equation}
then equating
\begin{equation}
    \mathcal{L}_\varepsilon = \sum_{k,k'; D_k + D_{k'} \neq 0} \frac{2}{D_k + D_{k'}}\braket{k|\partial_\varepsilon \bar{\rho}|k'}\ket{k}\bra{k'}.
\end{equation}
To facilitate convergence we began our calculations by expressing $\bar{\rho}$ in the discrete basis of exponentially-weighted Laguerre polynomials $\ket{\phi_n(\omega_0,\bar{\tau})}$ defined according to Eqs. (\ref{eq_WLket}) and (\ref{eq_WLwf}). We show in the Supplemental Material that matrix elements in this basis are given by:
\begin{eqnarray} \label{eq_rhobar_nm}
    \braket{\phi_n|\bar{\rho}_\tau|\phi_m} =&& \frac{1}{\tau\bar{\tau}} \int \mathrm{d}\omega \, \Biggl\{ \frac{P_0(\omega)}{\omega^2 + \sfrac{\Gamma_+^2}{4}} \nonumber \\ && \times \left[\frac{\sfrac{\Gamma_-}{2}+i\omega}{\sfrac{\Gamma_+}{2}+i\omega}\right]^n \left[\frac{\sfrac{\Gamma_-}{2}-i\omega}{\sfrac{\Gamma_+}{2}-i\omega}\right]^m \Biggr\},
\end{eqnarray}
where
\begin{equation}
    \Gamma_\pm = \frac{1}{\tau} \pm \frac{1}{\bar{\tau}}.
\end{equation}
For certain choices of $P_0(\omega)$ the integral in Eq. (\ref{eq_rhobar_nm}) might be analytically calculable via complex contour integration. In any case, it can be readily calculated numerically upon specifying $P_0(\omega)$. For the ensuing calculations we considered Gaussian broadening with spectral width parameter $\sigma$ such that:
\begin{equation}
    P_0(\omega) = \frac{1}{\sqrt{2\pi\sigma^2}}e^{-\sfrac{\omega^2}{2\sigma^2}}.
\end{equation}
Comparison of $\sigma$ and $1/\bar{\tau}$ determines the relative importance of lifetime broadening vs. pure dephasing. Equivalently, the product $(\sigma \bar{\tau})$ specifies the purity (Fig. \ref{fig:purity}), 
\begin{equation}
    \gamma(\bar{\rho}_{\bar{\tau}}) = \text{Tr}\left(\bar{\rho}^2_{\bar{\tau}}\right) = \frac{1}{\sqrt{4\pi}} \int_{-\infty}^{\infty} \mathrm{d}\Omega \, \frac{e^{-\Omega^2/4}}{1 + (\sigma \bar{\tau} \Omega)^2},
\end{equation}
of the limiting state
\begin{equation}
    \bar{\rho}_{\bar{\tau}} = \lim_{\varepsilon \to 1}\bar{\rho}.
\end{equation}
In the limit $\sigma \ll 1/\bar{\tau}$ we expect lifetime broadening to dominate and for the problem to revert to that of resolving $\tau_0$ and $\tau_1$ given the state in Eq. (\ref{eq_rho_mixed_old}) such that $\mathcal{K}_\varepsilon = \mathcal{K}_\varepsilon^\text{(max)}$. In the limit $\sigma \gg 1/\bar{\tau}$ pure dephasing dominates and we expect $\mathcal{K}_\varepsilon=\mathcal{J}_\varepsilon^\text{(TCSPC)}$, i.e. the QFI should asymptotically approach the CFI for TCSPC. This fact can be appreciated by inspection of the matrix elements of $\bar{\rho_\tau}$ in the temporal mode basis:
\begin{equation} \label{eq_rhobar_ttprime}
    \braket{t|\bar{\rho_\tau}|t'} = \frac{H(t)H(t')}{\tau}e^{-(t+t')/2\tau}e^{-i\omega_0(t-t')}e^{-(t-t')^2\sigma^2/2}.
\end{equation}
The effect of fixing $\tau$ and taking $\sigma \to \infty$ in Eq. (\ref{eq_rhobar_ttprime}) is to kill the off-diagonal elements of the matrix, leaving only populations which coincide with the outcome probability density of a TCSPC measurement.

\begin{figure}
    \centering
    \includegraphics{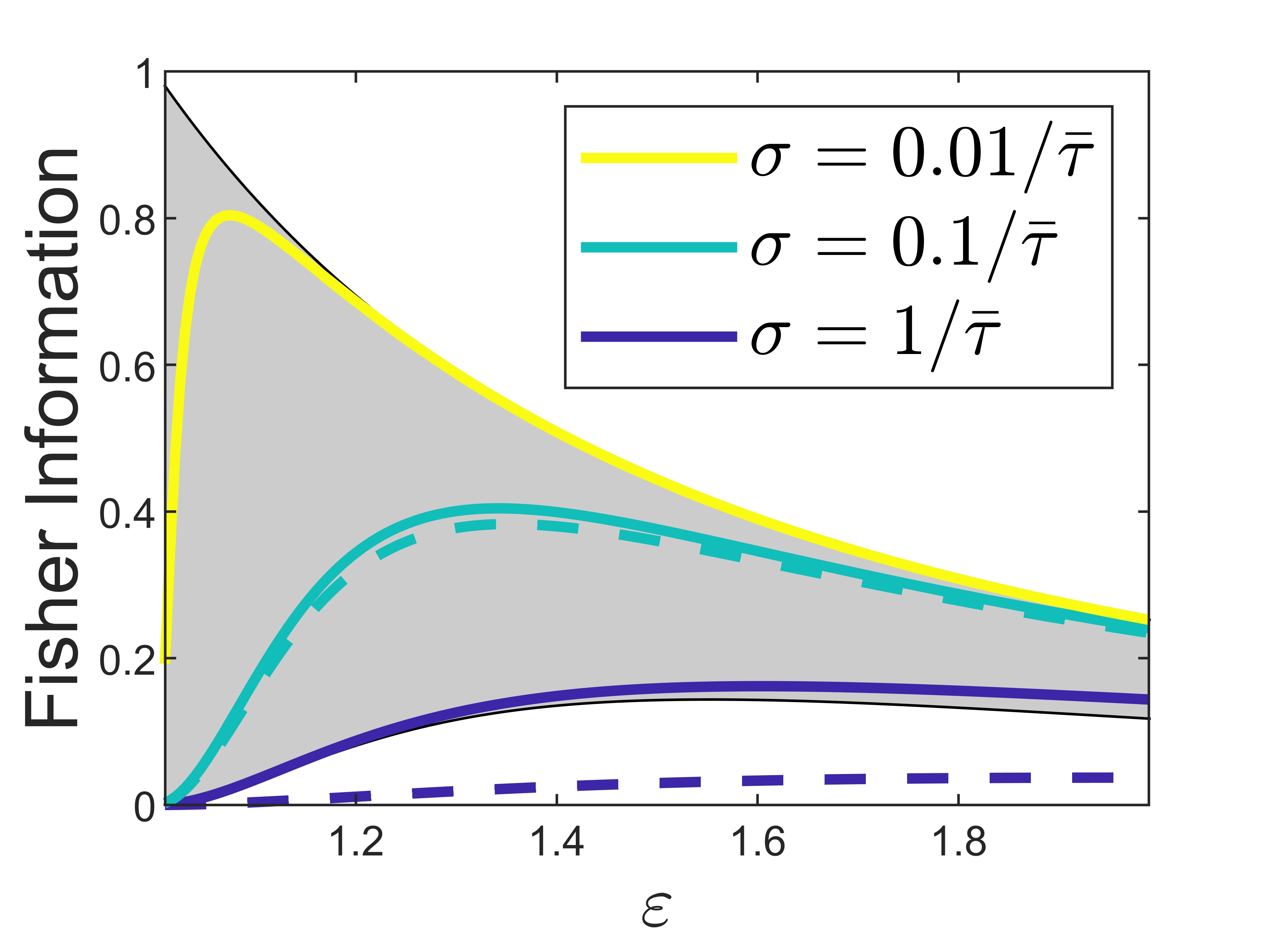}
    \caption{Fisher information associated with estimation of $\varepsilon$. The gray region is bounded above by the QFI in the limit $P_0(\omega) \to \delta(\omega)$ and below by the CFI associated with TCSPC. Fisher information which falls within the gray region therefore indicates a potential advantage over TCSPC. Solid colored lines demarcate the calculated QFI at varying degrees of spectral purity. Dashed colored lines indicate the CFI associated with projective measurement onto a set of WL modes as described in the text. The yellow dashed line is obscured as it overlaps almost completely with the yellow solid line.}
    \label{fig:FI comparison linear}
\end{figure}

\begin{figure}
    \centering
    \includegraphics{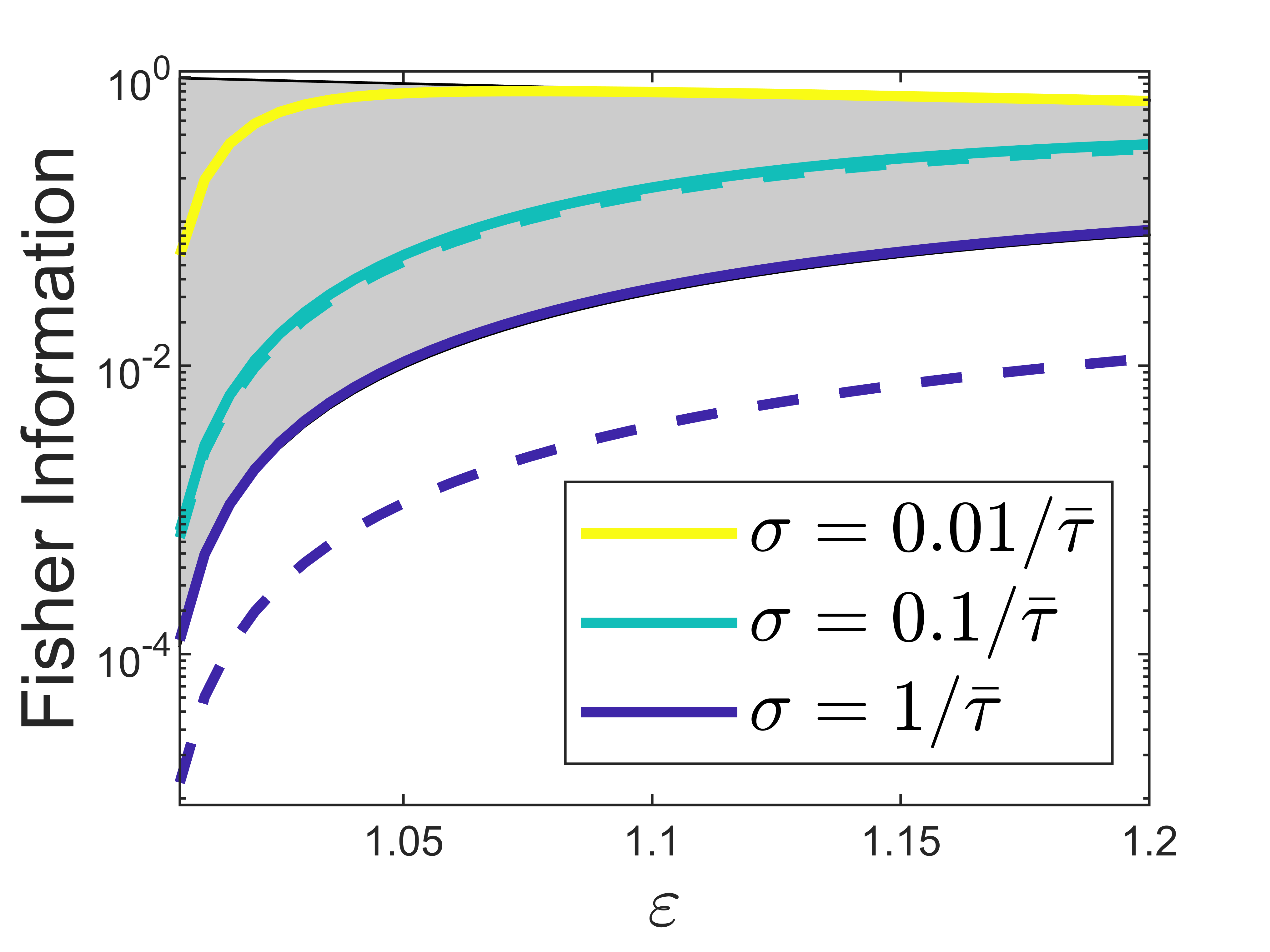}
    \caption{The same data as in Fig. \ref{fig:FI comparison linear} presented on a semilogarithmic scale.}
    \label{fig:FI comparison semilog}
\end{figure}

Figure \ref{fig:FI comparison linear} shows computed values of $\mathcal{K}_\varepsilon$ for $\sigma = 0.01/\bar{\tau}$, $0.1/\bar{\tau}$, and $1/\bar{\tau}$ (solid lines). The gray shaded region is bounded above by $\mathcal{K}_\varepsilon^\text{(max)}$ and below by $\mathcal{J}_\varepsilon^\text{(TCSPC)}$. For $\sigma \ll 1/\bar{\tau}$ we see that indeed $\mathcal{K}_\varepsilon \approx \mathcal{K}_\varepsilon^\text{(max)}$ over most of the domain, but for $\varepsilon$ sufficiently close to 1 the QFI begins to trend back down toward zero, indicating the resurgence of Rayleigh's Curse. Figure \ref{fig:FI comparison semilog} displays the same data as in Fig. \ref{fig:FI comparison linear} on a semilogarithmic scale. Close inspection reveals that despite the resurgence of Rayleigh's Curse, orders-of-magnitude resolution enhancement over TCSPC remains possible at $\varepsilon$ close to 1 and $\sigma < 0.1/\bar{\tau}$ [$\gamma(\bar{\rho}_{\bar{\tau}}) \gtrapprox 0.98$].

The color-coded dashed lines in Figs. \ref{fig:FI comparison linear} and \ref{fig:FI comparison semilog} mark the calculated CFIs, $\mathcal{J}_\varepsilon^\text{(WL)}$, associated with a projective measurement onto weighted Laguerre modes $\left\{\ket{\phi_n(\omega_0,\bar{\tau})}\bra{\phi_n(\omega_0,\bar{\tau})}\right\}_n$ truncated at $n_\text{max} = 100$. Actually only the single mode corresponding to $n=1$ is required to recover $>87\%$ and $>99\%$ 
of the available information near $\varepsilon = 1.05$ for $\sigma = 0.01/\bar{\tau}$ and $\sigma = 0.1/\bar{\tau}$, respectively. For $\sigma = 1/\bar{\tau}$ a projective measurement onto the first 100 WL modes is evidently far from optimal, as the dark blue dashed line falls well below the gray region. For $\sigma \geq 1/\bar{\tau}$ TCSPC does well to recover the available information; in this case, the most obvious measurement is the correct one.
\begin{figure}
    \centering
    \includegraphics{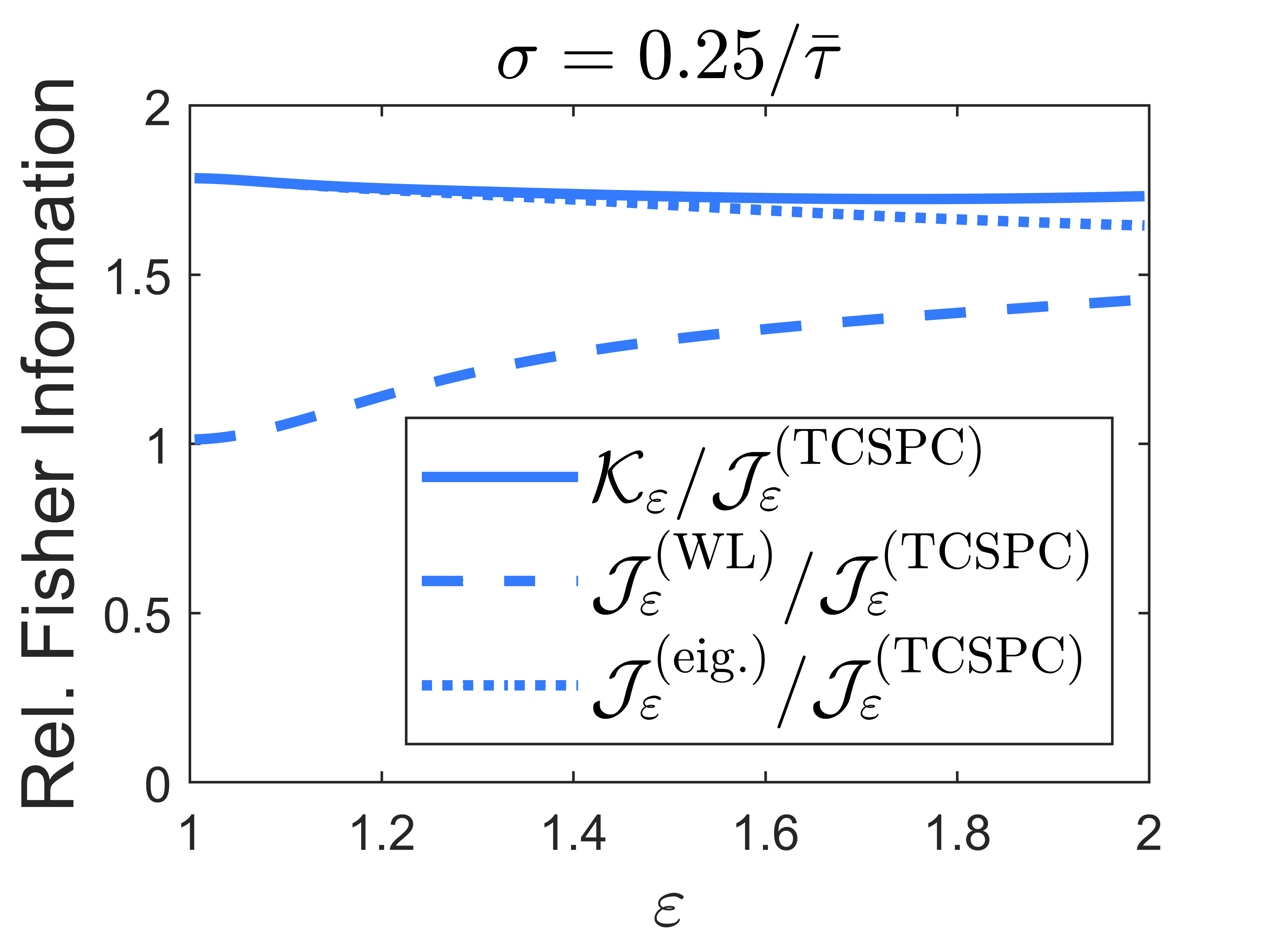}
    \caption{QFI (solid), CFI for WL projection (dashed), and CFI for optimal measurement near $\varepsilon=1$ (dotted) associated with the borderline case $\sigma = 0.25/\bar{\tau}$. Values are scaled by CFI of TCSPC.}
    \label{fig:special zoomed FI comparison}
\end{figure}
Figure \ref{fig:special zoomed FI comparison} depicts similar data for the borderline case of $\sigma = 0.25/\bar{\tau}$ [$\gamma(\bar{\rho}_{\bar{\tau}}) \approx 0.905$]. Here we scale the FIs by $\mathcal{J}_\varepsilon^\text{(TCSPC)}$. A modest information gain just under $2\times$ is available in this case, but it is not recovered by a measurement in the WL basis. For estimation of the single parameter $\varepsilon$, an optimal measurement can be constructed by projection onto the eigenstates of $\mathcal{L}_\varepsilon$. We calculate the performance of such a measurement for one choice of $\varepsilon$ by numerically diagonalizing $\mathcal{L}_\varepsilon$ after expressing $\bar{\rho}$ in the WL basis up to $n_\text{max} = 100$ (dotted line in Fig. \ref{fig:special zoomed FI comparison}).

The preceding analysis puts a finer point on exactly what quantum feature is needed to significantly surpass the resolution performance of TCSPC-- namely that coherences in the temporal mode basis must be preserved. Maximal information gain is possible in the idealized case that $\sigma \ll 1/\bar{\tau}$, for which subsequently collected photons are otherwise indistinguishable in the limit $\varepsilon \to 1$. 
\begin{figure}
    \centering
    \includegraphics{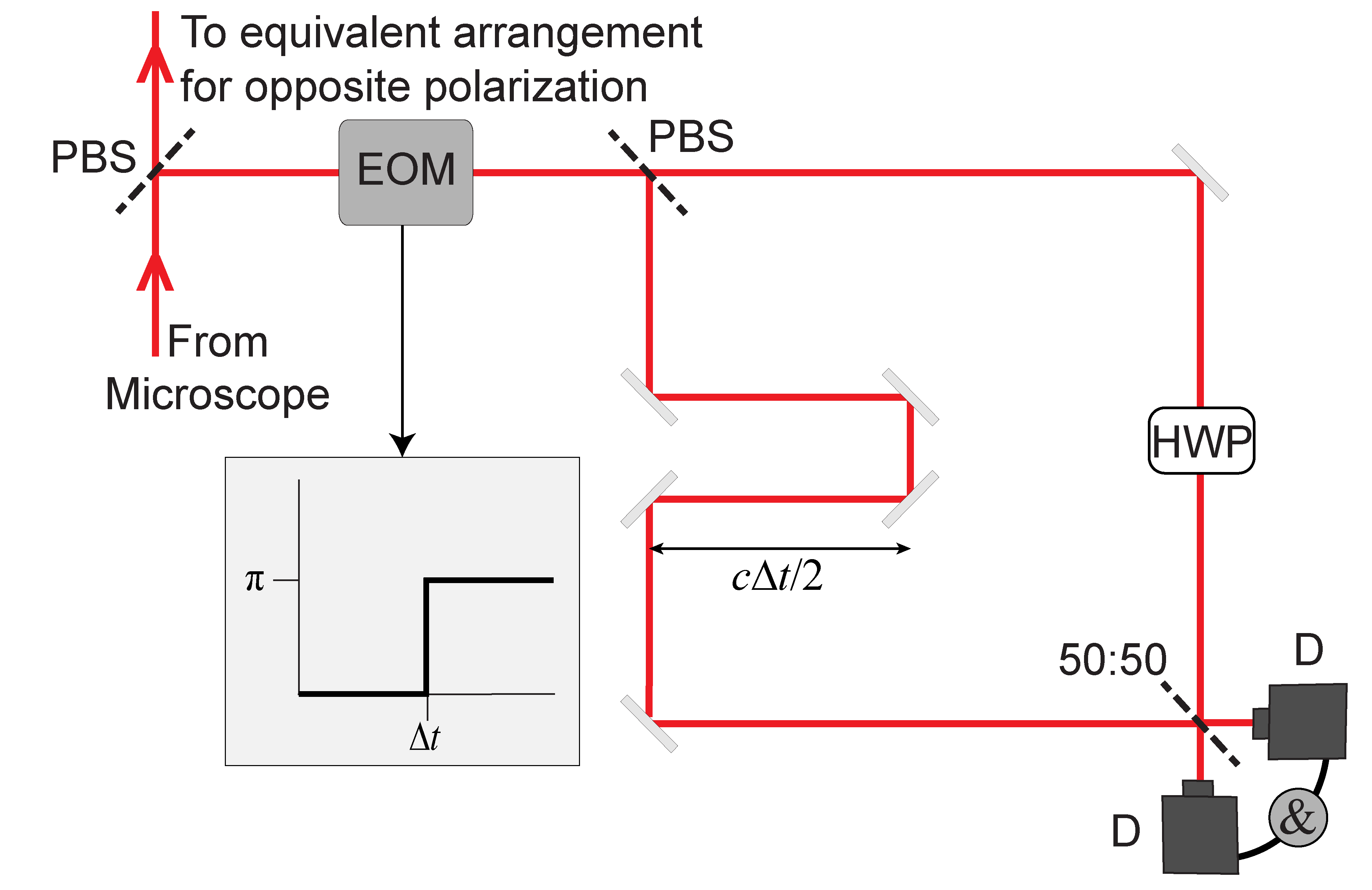}
    \caption{Proposed setup for resolving lifetimes via two-photon interferometry, including polarizing beam splitter (PBS), electro-optic modulator (EOM), half wave plate (HWP), non-polarizing 50:50 beam splitter, and coincidence detectors (D). The portion depicted here modulates the component of the emission that is s-polarized with respect to the first PBS. An analogous setup can be built on the other side to measure the p-polarized component.}
    \label{fig_HOMsetup}
\end{figure}
It's therefore apparent that in the limit $\sigma \to 0$ one has a choice between performing a tailored one-photon measurement as described in Ref. \cite{MitchellPRA2022} and a multi-photon interferometry measurement that exploits their near-indistinguishability. Having recognized this, we next analyze the lifetime resolving power of a two-photon Hong-Ou-Mandel type measurement, which has been shown previously to provide certain advantages in the context of the spatial resolution problem \cite{ParniakPRL2018}. We consider a hypothetical experiment using the apparatus depicted in Fig. \ref{fig_HOMsetup}. Subsequent excitation pulses are separated in time by an interval $\Delta t$ significantly longer than $\bar{\tau}$ such that emission in the window between the two pulses and the window after the second pulse is uncorrelated. We begin with a two-photon state of the form $\rho^{(2)} = \rho \otimes \rho$, where $\rho$ is the one-photon state defined in Eq. (\ref{eq_rho_mixed_old}). The QFI associated with $\varepsilon$ for this product state is simply twice that of $\rho$, i.e. $\mathcal{K}_\varepsilon^{(2)} = 2\mathcal{K}_\varepsilon^\text{(max)}$. The first collected photon is sent along one path and the second is sent along another by implementation of a switch synced with the second excitation pulse. This could be achieved, for example, by digitally switching an electro-optic modulator just before a polarizing beam splitter \cite{bowman2019electro}. The path of the first collected photon contains a delay stage to compensate the interpulse duration $\Delta t$. The path of the second collected photon contains a half wave plate to rotate the polarization to match that of the first photon. Then the two photons are brought together at either input port of a 50:50 beam splitter. There are four equally probable possibilities for the pair of lifetimes, which we denote $(\tau_0,\tau_0)$, $(\tau_0,\tau_1)$, $(\tau_1,\tau_0)$, and $(\tau_1,\tau_1)$. If we have $(\tau_0,\tau_0)$ or $(\tau_1,\tau_1)$ then the two photons are indistinguishable, and they will either both exit via port 1 or both via port 2. If instead we draw $(\tau_0,\tau_1)$ or $(\tau_1,\tau_0)$, there arises a small $\varepsilon$-dependent probability that the two photons emerge from opposite exit ports. By repeating the experiment and counting coincidences one can generate an estimate of $\varepsilon$. Our analysis detailed in the Supplemental Material shows that this measurement scheme recovers half of the available information, i.e. $\mathcal{J}_\varepsilon^\text{(\&)} = \mathcal{K}_\varepsilon^{(2)}/2$, when $\varepsilon \to 1$. Rayleigh's Curse is successfully averted. We can also conclude that given the choice between an optimal one-photon measurement and the described two-photon coincidence measurement, the former is superior in terms of information per photon detected. If, on the other hand, one has a choice between the two-photon coincidence measurement and a suboptimal one-photon measurement scheme that only recovers a fraction $\xi<1$ of the available information per photon, then which scheme is superior depends on whether $\xi$ is greater or less than 1/2. 

To this point in the analysis we have assumed that one photon is certainly collected within the interval immediately following each excitation pulse, and that it is not lost along its way to detection. In practice, a photon will only be successfully collected and relayed to the detector during this interval with some probability $p$, and under realistic conditions it is likely that $p \ll 1$. The relevant state of the field during this interval is then:
\begin{equation}
    \rho' = (1-p)\ket{\text{vac}}\bra{\text{vac}} + p\rho.
\end{equation}
The collective state of two such intervals, $\rho' \otimes \rho'$, contains two photons only with probability $p^2$. In this case a suboptimal one-photon measurement with efficiency $\xi$ is superior to the two-photon coincidence scheme if $\xi > p/2$. On the other hand, the HOM scheme offers a distinct potential advantage in that it does not require prior knowledge of the mean lifetime $\bar{\tau}$ to achieve super-resolution. By contrast, the correct choice of WL basis for an optimal one-photon measurement depends explicitly on $\bar{\tau}$, as estimated, e.g., from a preliminary TCSPC measurement.
\section{Conclusion}
In conclusion, we have effectively tightened the quantum bounds associated with resolution of optical spontaneous emission lifetimes by incorporating pure dephasing contributions to the spectral linewidth. When lifetime broadening dominates, a significant information gain can be uncovered by an appropriately tailored one- or two-photon measurement. When pure dephasing dominates, the conventional TCSPC measurement cannot be beat. It appears that any finite degree of pure dephasing causes the resolution QFI to scale back to zero for $\varepsilon$ sufficiently close to 1, indicating the eventual resurgence of the lifetime-analog of Rayleigh's Curse.
\begin{acknowledgements}
We thank Y. Wang, S. Bogdanov, and E. Goldschmidt for helpful comments on the manuscript. This work was supported via startup funds provided by the Department of Chemistry and the School of Chemical Sciences at the University of Illinois at Urbana-Champaign.
\end{acknowledgements}
\bibliography{mybib}

\end{document}